 \documentclass[final,5p,times,twocolumn]{elsarticle}

\usepackage{amssymb}
 \usepackage{amsthm}

 \usepackage{lineno}

\usepackage{graphicx}
\usepackage{dcolumn}
\usepackage{bm}
\usepackage{hyperref}
\usepackage{amsmath}
\usepackage{amsfonts}
\usepackage{sidecap}
\usepackage{float}
\usepackage{parskip}
\usepackage{grffile}

\usepackage{color}
\usepackage{hhline}

\makeatletter
\def\@eqnnum{{\normalsize \normalcolor (\theequation)}}
 \makeatother

\journal{Chaos, Solitons \& Fractals}

\begin{document}

\begin{frontmatter}



\title{Non-identical multiplexing promotes chimera states}


\author[label1]{Saptarshi Ghosh}
\ead{sapta15@gmail.com}
\author[label2]{Anna Zakharova}
\author[label1]{Sarika Jalan}
\ead{sarikajalan9@gmail.com}
\address[label1]{Complex Systems Lab, Discipline of Physics, Indian Institute of Technology Indore, Simrol, Indore 453552}
\address[label2]{Institut f{\"u}r Theoretische Physik, Technische Universit\"at Berlin, Hardenbergstra\ss{}e 36, 10623 Berlin, Germany}

\begin{abstract}
We present the emergence of chimeras, a state referring to coexistence of partly coherent, partly incoherent dynamics in networks of identical oscillators, in a multiplex network consisting of two non-identical layers which are interconnected. We demonstrate that the parameter range displaying the chimera state in the homogeneous first layer of the multiplex networks can be tuned by changing the link density or connection architecture of the same nodes in the second layer. We focus on the impact of the interconnected second layer on the enlargement or shrinking of the coupling regime for which chimeras are displayed in the homogeneous first layer. We find that a denser homogeneous second layer promotes chimera in a sparse first layer, where chimeras do not occur in isolation. Furthermore, while a dense connection density is required for the second layer if it is homogeneous, this is not true if the second layer is inhomogeneous. We demonstrate  that a sparse inhomogeneous second layer which is common in real-world complex systems, can promote chimera states in a sparse homogeneous first layer.
\end{abstract}


\end{frontmatter}


\section{Introduction.}
Coupled nonlinear systems can exhibit a plethora of complex interaction patterns and novel emergent phenomena, among which synchronization has been studied extensively due to its wide applicability to diverse fields in science and engineering \cite{rev.sync}.
Among various synchronization patterns, the chimera state representing a special type of partial synchronization has recently attracted considerable attention.
 This intriguing dynamical state is a hybrid state in which coherent and incoherent dynamics coexist in a coupled network of identical oscillators. Since their discovery in networks of phase oscillators \cite{chim.def}, chimera states have been reported for a variety of different systems including neural networks, time varying networks, planar oscillators, Boolean networks, Van der Pol oscillators, and for time-discrete maps as well as for time-continuous dynamical systems \cite{chim.def, chim.discrete.cont,chim.review}. Chimera states have recently been extended to quantum oscillators \cite{chim.quantum}. Although chimeras were first reported for non-local, non-global coupling \cite{chim.def}, recently chimera states have also been demonstrated for locally \cite{chim.local_coup} as well as globally \cite{chim.global} coupled networks.
There have been persistent efforts to gain a deeper insight into analyzing and controlling \cite{SIE14c,BIC15,OME16} chimera states. Experimentally, chimeras have been demonstrated for optical, chemical, mechanical, electronic and electro-chemical oscillators, 1D superconducting metamaterials etc \cite{chim.review}.
\begin{figure}[t]
 \centerline{\includegraphics[width=2.6in, height=1.0in]{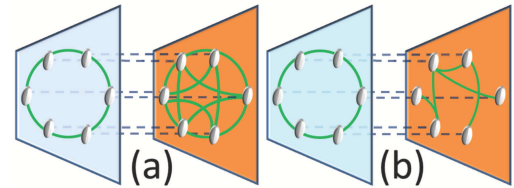}}
\vspace{-0.4cm}
 \caption{Schematic diagram depicting a multiplex network consisting of a 1D lattice and
(a) 1D lattice with different node degree (different value of nonlocal coupling range) and
(b) random network, respectively. We use this multiplex architecture as modeled by Eq.~(\ref{mul_mat}).}
 \label{fig1.schm_multiplex}
 \end{figure}

Furthermore, the investigation of multiplex networks has provided a new dimension to complex systems research, where chimeras have also been found \cite{chim.multiplex}. Multiplex networks (Fig.~\ref{fig1.schm_multiplex}) are defined as a collection of two or more layers which share the same nodes but have different connectivities in each layer \cite{mul_def}. Thus they describe networks which possess more than one type of interaction within the same elements, such as transport networks (with different means of travel as different layers), biochemical networks (with different signaling channels representing different layers), etc. \cite{mul_def}.

We investigate the emergence of chimera states in a homogeneous network of identical elements which is multiplexed with networks not necessarily having identical coupling environment. We refer to the network (layer) possessing nodes with identical coupling architecture as homogeneous network (layer).
We consider a 1D lattice with periodic boundary conditions (ring) for the homogeneous network. We show that the range of the coupling strength displaying chimera states in the homogeneous first layer can be controlled by changing the connection density and the coupling architecture of the second layer. We particularly consider two cases, (i) a multiplex network having two homogeneous layers with different connectivities, (ii) a multiplex network consisting of one homogeneous and one inhomogeneous layer.
\begin{figure}[t]
 \centerline{\includegraphics[width=2.2in,height=2.0in]{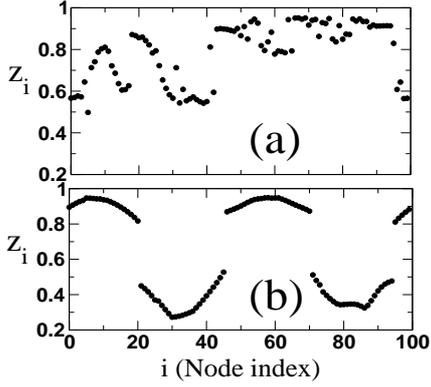}}
 \caption{Snapshots of the first layer for (a) isolated 1D ($\langle k \rangle =30$) and (b) 1D-1D ($\langle k^{(1)} \rangle=30$, $\langle k^{(2)} \rangle = 64$) multiplex network. Other parameters: $\varepsilon=0.33$ and $N^{(1)}=N^{(2)}=100$.}
 \label{fig1.2}
 \end{figure}

\section{Model.}
A network is a set of $N$ nodes and edges which can be represented
by an adjacency matrix $A$ of dimension $N\times N$ such that $A_{ij}=1$ if the nodes $i$ and $j$ are connected and $0$ if they are not. The adjacency matrix of two-layer multiplex networks can be expressed as
\begin{equation}
   A=
      \begin{pmatrix} A^{(1)} & I \\ I & A^{(2)} \end{pmatrix},
\label{mul_mat}
\end{equation}
where $A^{(1)}$ ($A^{(2)}$) represents the adjacency matrix of the first (second) layer and $I$ is an $N \times N$ identity matrix representing the interactions between the two layers. The number of nodes is the same in both layers.
In the following, we investigate the dynamical evolution of one layer ($A^{(1)}$)
with a fixed, homogeneous coupling scheme in dependence on the link density and the architecture of
the other layer ($A^{(2)}$).

We consider a discrete-time map $z_i(t+1)=f(z_i(t))$, $z_i(t)\in \mathbb{R}, i=1,...,2N$ as a real dynamical variable at time $t$ for the $i^{th}$ node. We further integrate the underlying network topology as \cite{multi_appl}
\begin{equation}
z_i(t+1)=f(z_i(t))+\frac{\varepsilon}{(k_i + 1)} \sum_{j=1}^{2N} A_{ij}[ f(z_j(t))-f(z_i(t)) ]
\label{eq.evol}
\end{equation}
where $ k_{i}$ = $\sum_{j=1}^{N}A_{ij}$ is the degree of the $i^{th}$ node in its own layer and $\varepsilon$ is the overall coupling constant, assuming $0\le \varepsilon \le1$. Furthermore the average degree (node degree) of the network is defined as $\langle k \rangle = \sum_{i=1}^{N} k_{i}/N$, where $N$ represents number of nodes in the network.
\begin{figure*}[t]
\centerline{\includegraphics[width=4.8in, height=2.0in]{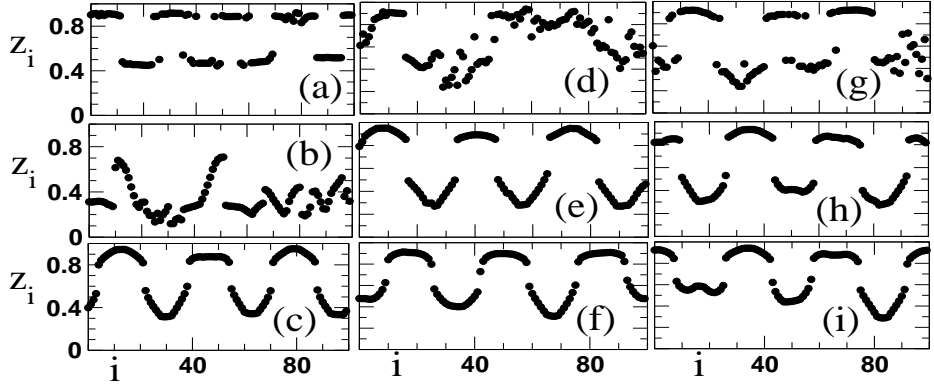}}
\caption{Snapshots of the first layer for a 1D-1D multiplex network for (a,d,g) $\varepsilon=0.33$, (b,e,h) $\varepsilon=0.38$ and (c,f,i) $\varepsilon=0.44$. The node degrees of the first and second layer are  $\langle k^{(1)}\rangle=20$ and (a,b,c) $\langle k^{(2)} \rangle=10$, (d,e,f) $\langle k^{(2)} \rangle=40$, (g,h,i) $\langle k^{(2)} \rangle=64$, respectively. Other parameters: $N^{(1)}=N^{(2)}=100$}
 \label{fig2.2}
 \end{figure*}
We choose the logistic map $f(z)=\mu z (1-z)$ in the chaotic regime ($\mu=4.0$) as local dynamics. The coupled logistic maps display rich dynamical behavior in this regime \cite{logsMap} and has been widely used as a paradigmatic model for various complex dynamical phenomena in real-world systems \cite{logsMap_appl}. One such phenomenon is the emergence of hybrid patterns in the form of chimera states \cite{chimera_logs}.

The dynamical state of the network (represented by $z_i(t)$) is defined as coherent \cite{chim.discrete.cont} if
\begin{equation}
\lim\limits_{N \rightarrow \infty} \lim\limits_{t \rightarrow \infty}\sup\limits_{i,j  \in U_{\xi}^N (x)} \mid{z_i(t)-z_j(t)} \mid \rightarrow 0 \, \, \text{for} \, \, \xi\; \rightarrow 0
\label{eq.cohr}
\end{equation}
where $U_{\xi}^{N} (x) = \{ j : 0 \le j \le N, \mid{\frac{j}{N} - x} \mid <  \xi \}$ represents the network neighborhood of any point $x \in S^1$, i.e., of the 1D network. Geometrically, coherence means that in the continuum limit $N \to \infty$ snapshots of the state $z_i(t)$ approach a smooth profile $z(x, t)$.

Any discontinuity appearing in the profile reflects the spatial incoherence. Further, to clearly identify chimera states, we employ a normalized probability distribution function $g(\mid{\bar{D}}\mid)$ of the Laplacian distance measure $|\bar{D}(t)|$ defined in \cite{chim.class}, and the correlation measure
\begin{equation}
g_0(t)= \int_{0}^{\delta} g (\mid{\bar{D}(t)}\mid)  d(\mid{\bar{D}(t)}\mid)
\label{eq.dmsr}
\end{equation}
where $|\bar{D}(t)|$ is a vector with components $d_i(t)$ defined as $d_{i}(t)=| (z_{i+1}(t)-z_{i}(t))-(z_{i}(t)-z_{i-1}(t))|\}$ identifies the presence of strong local curvature in an otherwise smooth spatial profile \cite{suppl}. The upper limit $\delta$ (Eq.~\ref{eq.dmsr}) denotes a small positive threshold value. The $g_0(t)$ essentially measures the relative size of spatially coherent regions, and ideally an intermediate value between 0 and 1 indicates a chimera state~\cite{note1}. However, even an incoherent state may have a small portion of nodes which can cluster together spatially, leading to a non-zero value of $g_0$. We numerically find that $0.4\lesssim g_0 \lesssim 0.8$ provides a best estimation of the parameter regime displaying Chimera states. We highlight the region in the figures (in $g_0-\varepsilon$ plane) for an easy comprehension. Furthermore, discussion on special initial condition considered here can be found in Ref.\cite{note2}.

\section{Chimeras in multiplex networks with connectivity mismatch.}
A multiplex network consisting of exactly the same architecture (node degree $\langle k^{(1)} \rangle=\langle k^{(2)} \rangle$) in both layers has been reported to exhibit chimera states in Ref.~\cite{chim.multiplex} which is same as a single layer case \cite{chim.discrete.cont}. However, this kind of identical network architecture of both layers is hard to find in real-world systems. For example, a transportation network consists of layers representing different modes of travel. The air travel layer may be more sparsely connected than the rail or bus layer. Similarly, in a communication network like the internet the optical fiber layer will have a much sparser connectivity than the traditional cable network. Motivated by this, we investigate the behavior of chimera states for multiplex networks with nonidentical layers, possessing properties which are closer to those of real-world systems.

First, we consider the case of both layers being represented by a homogeneous nonlocal coupling topology but with a connectivity mismatch between the layers (Fig.~\ref{fig1.schm_multiplex}(a)). Specifically, we choose a nonlocal coupling range with $P^{(1)}$ and $P^{(2)}$ neighbors to each side in the two layers, respectively. This assignment corresponds to a constant node degree of $\langle k^{(1)} \rangle=2 P^{(1)}$ and $\langle k^{(2)} \rangle=2 P^{(2)}$, respectively.  We find that chimera states emerge in the sparse first layer, in contrast to the single layer case, when it is multiplexed with a dense second layer. Fig.~\ref{fig1.2}(a) shows that no chimera exists for the single layer network (incoherent state), and Fig.~\ref{fig1.2}(b) depicts a chimera in the same sparse layer upon multiplexing with a dense layer.  Furthermore, Fig.~\ref{fig2.2} display that the range of the $\varepsilon$ for which the chimera state exists in the first layer is enlarged as the second layer becomes denser. In Fig.~\ref{fig2.2} (left column), when the second layer has the node degree $\langle k^{(2)} \rangle =10$, a clear chimera state is only found for a very large value of $\varepsilon$ (Fig.~\ref{fig2.2} (c)), whereas for $\langle k^{(2)} \rangle =40$ (Fig.~\ref{fig2.2} (middle column)) the  chimera state exists for a larger range (Fig.~\ref{fig2.2} (e)-(f)). With further increasing node degree of the second layer, say for $\langle k^{(2)} \rangle =64$ (Fig.~\ref{fig2.2} (right column)), the chimera state in the sparser first layer exists for almost all $\varepsilon$ values as depicted in Fig.~\ref{fig2.2} (g)-(i).

So far we have kept the degree of the sparse first layer fixed and have varied the node degree of the second layer, demonstrating that with increasing node degree of the second layer chimeras occur for a larger range of $\varepsilon$. The same is true if we fix the node degree of the dense second layer and change the degree of the sparse first layer. Again, a stronger connectivity mismatch leads to a larger range of $\varepsilon$ for which chimeras are observed in the sparser first layer. Note that the dense second layer still exhibits chimeras in an intermediate $\varepsilon$ range (similar to the case where both layers of the multiplex network consist of dense regular coupling topology) regardless of the connection density of the sparse first layer \cite{suppl}.

\begin{figure}[b]
 \centerline{\includegraphics[width=0.9\columnwidth]{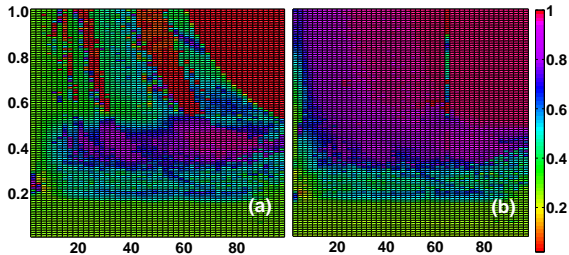}}
 \caption{(Color online) The normalized correlation measure $g_0$, calculated for layer 1, is plotted in the parameter plane ($\langle k^{(1)} \rangle, \varepsilon$) for (a) 1D-1D multiplex network  with the same node degree in both layers ($\langle k^{(1)} \rangle = \langle k^{(2)} \rangle = \langle k \rangle$), (b) mismatched 1D-1D multiplex network where the second layer has the fixed node degree $\langle k^{(2)} \rangle= 64$. Note the chimera tongue around $\langle k^{(1)} \rangle= 64$ in (b). Parameters: $N^{(1)}=N^{(2)}=100$ and $\delta=0.01(max(|D|))$ \cite{chim.class};  $g_0$ is averaged over $1000$ time steps.}
 \label{fig.4}
 \end{figure}

To present a comprehensive picture of multiplexing with a denser layer that promotes chimeras in a sparse network, we plot a diagram of the parameter regimes in the plane of the node degree $\langle k^{(1)} \rangle$ and $\varepsilon$. The density plot shows the correlation measure $g_0$ of the sparse first layer (Fig.~\ref{fig.4}). Fig.~\ref{fig.4} (a) depicts a network with the same node degree $\langle k \rangle$ in both layers. Now we keep the node degree of the dense second layer fixed and vary the node degree of the first layer from very sparse to very dense. Fig.~\ref{fig.4} (b) shows that there exists a regime of chimera states in the first layer ($0<g_0<1$) in the ($\langle k^{(1)} \rangle,\varepsilon$) parameter plane at intermediate values of $\varepsilon$ and $\langle k^{(1)} \rangle= 64$, which corresponds to two identical layers. The light-colored region to the very left of Fig.~\ref{fig.4} (b), corresponding to a sparse layer (low $\langle k^{(1)} \rangle$) multiplexed with a dense layer, also indicates chimera states in a parameter regime of large $\varepsilon$ where they are not found in a single layer (cf. $\langle k^{(1)} \rangle = \langle k^{(2)} \rangle$ in Fig.~\ref{fig.4} (a)). On the other hand, for large $\langle k^{(1)} \rangle$, where both layers are dense, chimeras are only found in a small range at intermediate values of $\varepsilon$  (light color in Fig.~\ref{fig.4} (b)).

\begin{figure}[t]
\centerline{\includegraphics[width=3.4in, height=1.3in]{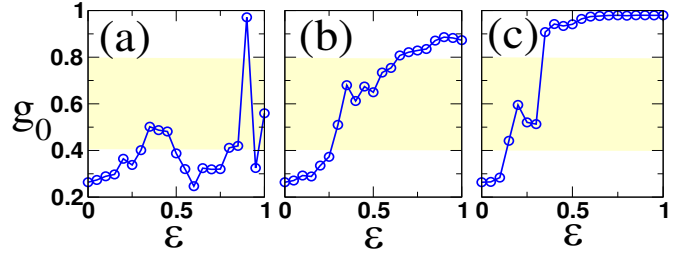}}
\caption{(Color online) Correlation measure $g_0$ vs $\varepsilon$ for the first layer of a mismatched 1D-1D multiplex network characterizing the chimera behavior in the sparse layer as a consequence of multiplexing.
Node degrees: (a) $\langle k^{(1)} \rangle= 10$,  $\langle k^{(2)} \rangle = 10$, (b) $\langle k^{(1)} \rangle= 10$, $\langle k^{(2)} \rangle = 64$, (c) $ \langle k^{(1)} \rangle = 64$, $ \langle k^{(2)} \rangle = 80$.
Other parameters as in Fig.~\ref{fig.4}. Highlighted area indicates parameter regime for chimera states.}
 \label{fig3.1D_1D_scatter}
 \end{figure}

To further illustrate this issue in Fig.~\ref{fig3.1D_1D_scatter}, we plot the correlation measure $g_0$ as a function
of $\varepsilon$  for a multiplex network with mismatched node degree of the two layers.
In panel (a) layers 1 and 2 are sparse, in (b) layer 2 is more densely connected than layer 1, and in panel (c) both layers are dense.
A multiplex network consisting of two sparse layers has a low value of the correlation measure $g_0$ indicating incoherent dynamics in layer 1 (Fig.~\ref{fig3.1D_1D_scatter}(a)). However, with increasing connectivity of layer 2, the critical coupling strength (i.e., $\varepsilon$ value for which the network dynamics exhibits a transition from the chimera to the completely coherent state) increases in layer 1, indicating an extended regime of chimeras. In fact, the sparse layer (layer 1) demonstrates absence of the completely coherent regime ($g_0\approx 1$) when multiplexed with a dense layer (Fig.~\ref{fig3.1D_1D_scatter}(b)). Furthermore, for a multiplex network consisting of two dense layers, both layers show a typical chimera regime in an intermediate range of $\varepsilon$ as exhibited by identical dense layers (Fig.~\ref{fig3.1D_1D_scatter}(c)). Thus, for multiplex networks with dense layers, the individual layers do not exhibit any change in the critical $\varepsilon$
value  for the occurrence of chimera states.

\section{Chimera patterns upon multiplexing with inhomogeneous network topology.}
Here, we discuss the impact of inhomogeneous network architecture in the second layer of a multiplex network on the
emergence of chimera patterns in the homogeneous first layer. We consider a multiplex network where a 1D ring lattice with homogeneous nonlocal coupling (layer 1) is multiplexed with an
inhomogeneous network having a random architecture (layer 2), see Fig.~\ref{fig1.schm_multiplex} (b).
Since the network architecture represented by the second layer does not consist of nodes which are ordered by nearest neighbor coupling configurations, it is not straightforward to define chimera states in the classical sense for the second layer. All the figures and discussions in the following correspond to the dynamics of the 1D lattice in the first layer.

\begin{figure}[t]
\centerline{\includegraphics[width=3.4in, height=1.3in]{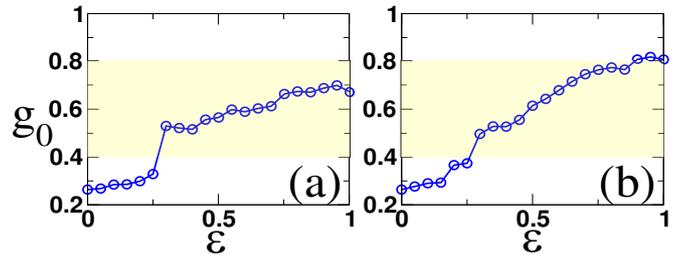}}
\caption{Correlation measure $g_0$ vs $\varepsilon$ for the homogeneous first layer of a (a) 1D-ER (b) 1D-SF multiplex network. Node degrees $\langle k^{(1)} \rangle = \langle k^{(2)} \rangle = 10$. Other parameters as in Fig.~\ref{fig.4}. Highlighted area indicates parameter regime for chimera states.}
\label{fig.9.1D_ER_SF.scatter}
\end{figure}

\begin{figure}[t]
 \centerline{\includegraphics[width=0.9\columnwidth]{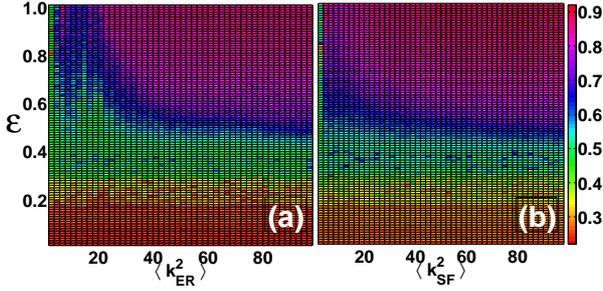}}
 \caption{(Color online) Map of regimes for a multiplex networks consisting of
one 1D layer and one inhomogeneous layer, where $g_0$ is calculated for layer 1:
(a) 1D-ER multiplex networks and (b) 1D-SF multiplex network. The first layer has
the node degree $\langle k^{(1)} \rangle= 10$. Other parameters as in Fig.~\ref{fig.4}.}
 \label{fig3.1D_er_sf_phase}
 \end{figure}

To construct the inhomogeneous layer, first we use an Erd\"{o}s-R\'{e}nyi (ER) network. We consider a
multiplex network consisting of a dense 1D lattice (layer 1) and an ER random network (layer 2).
The dense 1D layer exhibits chimera states at intermediate coupling values without any enhancement
(as compared to the single-layer case) regardless of the connection density of the inhomogeneous ER
layer \cite{suppl}. An interesting phenomenon, however, occurs when a sparse 1D layer is multiplexed with a random network. Unlike the case of multiplex networks consisting of two sparse homogeneous layers, if one layer is represented by a random connection architecture, for the same connection density the homogeneous layer exhibits chimeras. The sparse 1D layer which does not exhibit chimeras upon multiplexing with a sparse homogeneous layer (Fig.~\ref{fig3.1D_1D_scatter} (a)), starts displaying chimeras when multiplexed with a sparse inhomogeneous layer (Fig.~\ref{fig.9.1D_ER_SF.scatter} (a)-(b)). Moreover, multiplexing with a sparse ER networks is more favorable for the emergence of chimeras in the homogeneous layer than multiplexing with a dense layer. The critical $\varepsilon$ value in the sparse 1D lattice increases with decreasing average connection density $\langle k^2_{ER} \rangle$ of the ER layer (Fig.~\ref{fig3.1D_er_sf_phase}).
The chimera regime expands as the ER layer becomes  sparser. Fig.~\ref{fig3.1D_er_sf_phase}(a)
depicts a larger range of $\varepsilon$ for which chimeras are observed in the 1D layer due to its
multiplexing with a sparser random layer.
One further point to be noted is that for multiplex networks consisting of two homogeneous layers,
larger average connectivity is more favorable for synchronization in one layer, however,
for multiplex networks consisting of one homogeneous and one inhomogeneous (say ER) layer,
enhancement in average connectivity of inhomogeneous layer leads to a shrinking coupling range
for which chimeras are observed. Nevertheless, for all combinations of average degree,
multiplexing with inhomogeneous layers yields a larger coupling range for chimeras than multiplexing with a homogeneous layer. Furthermore, to demonstrate the robustness of (i) the emergence of
chimera states in a sparse 1D layer upon multiplexing with an inhomogeneous layer,
and (ii) shrinking of the range of $\varepsilon$ for which chimeras are observed
in a sparse 1D layer with increasing connection density of the inhomogeneous region,
we consider a multiplex network consisting of a 1D and a scale-free (SF) layer.
The SF network is generated using the preferential attachment model \cite{sf_network}. For this arrangement as well,
chimera states emerge in the sparse homogeneous layer upon multiplexing with another sparse SF network.
Additionally, an increase in the average connectivity of the SF network $\langle k^2_{SF} \rangle$
yields a similar shrinking of the chimera regime in the sparse
1D layer (Fig.~\ref{fig3.1D_er_sf_phase}(b)). Furthermore, similar to the dense 1D-ER multiplex network, a 1D-SF multiplex network consisting of a dense 1D layer does not show any enhancement or suppression of the chimera state occurring in the 1D layer as compared to the corresponding isolated 1D network
regardless of the connection density of the layer that it is multiplexed with \cite{suppl}.

The reason behind the emergence of chimera states upon multiplexing with an inhomogeneous layer
in those sparse networks which do not exhibit chimeras upon multiplexing with a homogeneous
layer of the same average connectivity, seems to lie
in the existence of high degree nodes in the inhomogeneous layer. To obtain high degree nodes
in a layer of a 1D-1D multiplex network, one needs to enhance the average connectivity
of the layer. Hence, we find chimeras for the sparse 1D - dense 1D multiplex network and do not observe
chimeras when both homogeneous layers are sparse. Whereas for a 1D-ER multiplex network,
even if both layers are sparse, there may be a large mismatch in the degrees of a
few pairs of mirror nodes.  For 1D-SF multiplex networks, multiplexing has a more pronounced effect
which may arise from a higher degree mismatch for a few pairs of mirror nodes due to the existence of hub
nodes in the SF layers (Fig.~\ref{fig.9.1D_ER_SF.scatter}).

\section{Conclusion.}
To summarize, we have shown that the occurrence of chimera states in a layer of
multiplex network depends not only on the coupling strength or the initial condition but also on the network architecture of the layers that it is multiplexed with.
Multiplex networks with non-identical layers promote the appearance of chimeras in a sparse
homogeneous layer as compared to the case of single-layer networks as well as to the case of multiplex network consisting of identical sparse homogeneous layers. The average connectivity (mean node degree) of one
layer plays a crucial role in governing the appearance of chimera patterns in the other layer.
Our investigations reveal that by controlling the mean node degree of one layer in the multiplex
network, one can tune the coupling strength for which chimeras are observed in the other layer.
Furthermore, the behavior of chimeras in the layer with homogeneous coupling depends on the architecture of the other layers in multiplex networks. If both layers are homogeneous, multiplexing with a denser second layer promotes the occurrence of chimeras in the sparse first layer, whereas, if multiplexing is done with an inhomogeneous layer,
it enhances the parameter range for the appearance of
chimera states in the homogeneous layer even if both layers are sparse.
The emergence of chimeras in networks upon multiplexing with an
inhomogeneous layer as well as enhancement of the coupling range for which chimeras appear in the sparse
layer  is more prominent if multiplexing is done with a scalefree network.
Recently, it has been pointed out that chimera states may have promising applications in
understanding various complex processes in nature including epileptic seizures \cite{chim.epileptic}, human or mammal uni-hemispheric sleep \cite{chim.TAM16}, motion of heart vessels for ventricular fibrillation
\cite{chim.ventricular} and in ecology \cite{chim.ecological_network}.  The results presented in this paper may help us to
gain deeper insight into the emergence and impact of chimera states in real-world networks which inherently possess a multi-layer architecture.

\section{acknowledgments}
SJ acknowledges DST project grant (EMR/2014/000368 and EMR/2016/001921) for financial support. SG acknowledge DST Government of India for the INSPIRE fellowship (IF150149) and members of Complex Systems Lab for useful discussions. AZ acknowledge support from DFG in the framework of SFB 910. SJ thanks hospitality and support of Max-Planck-Institute for the Physics of Complex Systems, Dresden where this work was completed. We acknowledge Anil Kumar and Prof. Eckehard Sch\"{o}ll for valuable discussions during initial phase of the work.


\begin{thebibliography}{99}
\bibitem{note1} The Laplacian distance measure $|\bar{D}(t)| = \{d_i(t): d_{i}(t)=| (z_{i+1}(t)-z_{i}(t))-(z_{i}(t)-z_{i-1}(t))|\}$ (Eq.~\ref{eq.dmsr}) captures the spatial distance between the neighboring nodes. The nodes belonging to the incoherent region posses high value of the laplacian distance measure owing to the high local curvature in the spatial profile \cite{suppl}. The normalized probability distribution function $g (\mid{\bar{D}}\mid)$ (Eq.~\ref{eq.dmsr}) categories the number of nodes having particular spatial distances. The correlation measure $g_0$ (Eq.~\ref{eq.dmsr}) calculates the relative amount of nodes having spatial distance below a certain threshold $\delta$ where $\delta$ (Eq.~\ref{eq.dmsr}) sets a limit below which the spatial curve is considered smooth indicating coherent nodes. The $\delta=0.01(max(|D|))$ \cite{chim.class} is taken such that it distinguishes between different dynamical states. So, the $g_0(t)$ indicates the relative size of spatially coherent regions.The measure $g_0(t)$ taking value $1(0)$ indicates all (none) of the nodes have very small spatial distance with respect to its neighbors and thus form a smooth (discontinuous) spatial profile. Any intermediate value of $g_0$ between 0 and 1 shows presence of both coherent and incoherent regions denoting a chimera state \cite{chim.class}.
\end{thebibliography}
\end{document}